\documentclass{article}
\usepackage{spconf,amsmath,graphicx}

\usepackage[utf8]{inputenc}
\usepackage[T1]{fontenc}
\usepackage[ngerman]{babel}

\usepackage{enumitem}
\setlist{nosep, leftmargin=14pt}
\usepackage[colorlinks=true, allcolors=blue]{hyperref}
\usepackage{mwe} 
\usepackage{multirow}
\usepackage{makecell}
\usepackage{adjustbox}
\usepackage{xcolor}
\usepackage{float}
\usepackage{subcaption}

\title{Automated classification of multi-parametric body MRI series}
\name{Boah Kim, Tejas Sudharshan Mathai, Kimberly Helm, Ronald M. Summers}
\address{Imaging Biomarkers and Computer-Aided Diagnosis Laboratory, \\ Radiology and Imaging Sciences, National Institutes of Health Clinical Center, Bethesda, MD, USA}
\begin{document}
%
\maketitle

\begin{abstract}

\noindent
Multi-parametric MRI (mpMRI) studies are widely available in clinical practice for the diagnosis of various diseases. As the volume of mpMRI exams increases yearly, there are concomitant inaccuracies that exist within the DICOM header fields of these exams. This precludes the use of the header information for the arrangement of the different series as part of the radiologist's hanging protocol, and clinician oversight is needed for correction. In this pilot work, we propose an automated framework to classify the type of 8 different series in mpMRI studies. We used 1,363 studies acquired by three Siemens scanners to train a DenseNet-121 model with 5-fold cross-validation. Then, we evaluated the performance of the DenseNet-121 ensemble on a held-out test set of 313 mpMRI studies. Our method achieved an average precision of 96.6\%, sensitivity of 96.6\%, specificity of 99.6\%, and $F_1$ score of 96.6\% for the MRI series classification task. To the best of our knowledge, we are the first to develop a method to classify the series type in mpMRI studies acquired at the level of the chest, abdomen, and pelvis. Our method has the capability for robust automation of hanging protocols in modern radiology practice.


\end{abstract}

\section{Introduction}

Multi-parametric MRI (mpMRI) is a widely used imaging technique to visualize the anatomy of the body including the chest, abdomen, and pelvis, and to diagnose various diseases such as prostate cancer \cite{Turkbey2023} and lymphadenopathy \cite{Mathai2023_LN}. Due to its widespread availability today, the volume of MRI exams has increased \cite{OECD2023}. During an MRI exam, a number of different series are acquired using a combination of echo times, repetition times, radio-frequency pulses, and other parameters. Some of the sequences provide complementary information related to disease status and are used by radiologists simultaneously. At the acquisition time, relevant information related to the series is usually stored in the ``Series Description'' and ``Protocol Name'' fields of the DICOM header.

These DICOM header fields are of particular importance as logic-based rules are applied to them. These rules facilitate the arrangement of the different series in the mpMRI study in the Picture Archival and Communication Systems (PACS) viewer. This arrangement constitutes the hanging protocol for the radiologist who reads the study. However, there are a variety of MRI scanners from different manufacturers, diverse MRI protocols employed at different institutions across the world, and subjective preferences of the MRI technologist. These factors affect the information entered into the DICOM fields. As described in previous works \cite{Gueld_2002, Anand2023_CTClass, Zhu2022, Baumgartner2023_prostateMRI}, $\sim$16\% of the DICOM headers contain fields that have heterogeneous, inconsistent, and subjective information. Such inaccuracies impede the use of the DICOM tags for automatic series categorization \cite{Gueld_2002}. Radiologists presently rearrange the series based on their preferences and read the exam. To ameliorate the radiologist oversight necessary, an automated approach to classify the MRI sequence type would be beneficial to improving their reading efficiency. 

Prior works in this area have mostly focused on classifying the series type in brain MRI studies \cite{Liang2021_metadataRF, Ranjbar2020, Mello2021_ResNet18, Noguchi2018, Remedios2018, Chou2022}. Liang et al. \cite{Liang2021_metadataRF} achieved near-perfect classification accuracy for brain MRI series identification. Also, the existing research targeted towards the same task in the body has focused on either the liver \cite{Zhu2022} or the prostate \cite{Baumgartner2023_prostateMRI}. Zhu et al. \cite{Zhu2022} developed a 3D pyramid pooling architecture (3DPP) for abdominal MRI series classification and achieved a ``strict'' accuracy of 90.4\% with their 3DPP InceptionV1 network. Baumgartner et al. \cite{Baumgartner2023_prostateMRI} used a ResNet-18 model to classify the type of prostate MRI sequence and obtained an overall accuracy of 99.8\%. To the best of our knowledge, there is no prior work on classifying the series types in chest, abdomen, and pelvis MRI studies.

\begin{figure*}[!t]
\centering
\includegraphics[width=\linewidth]{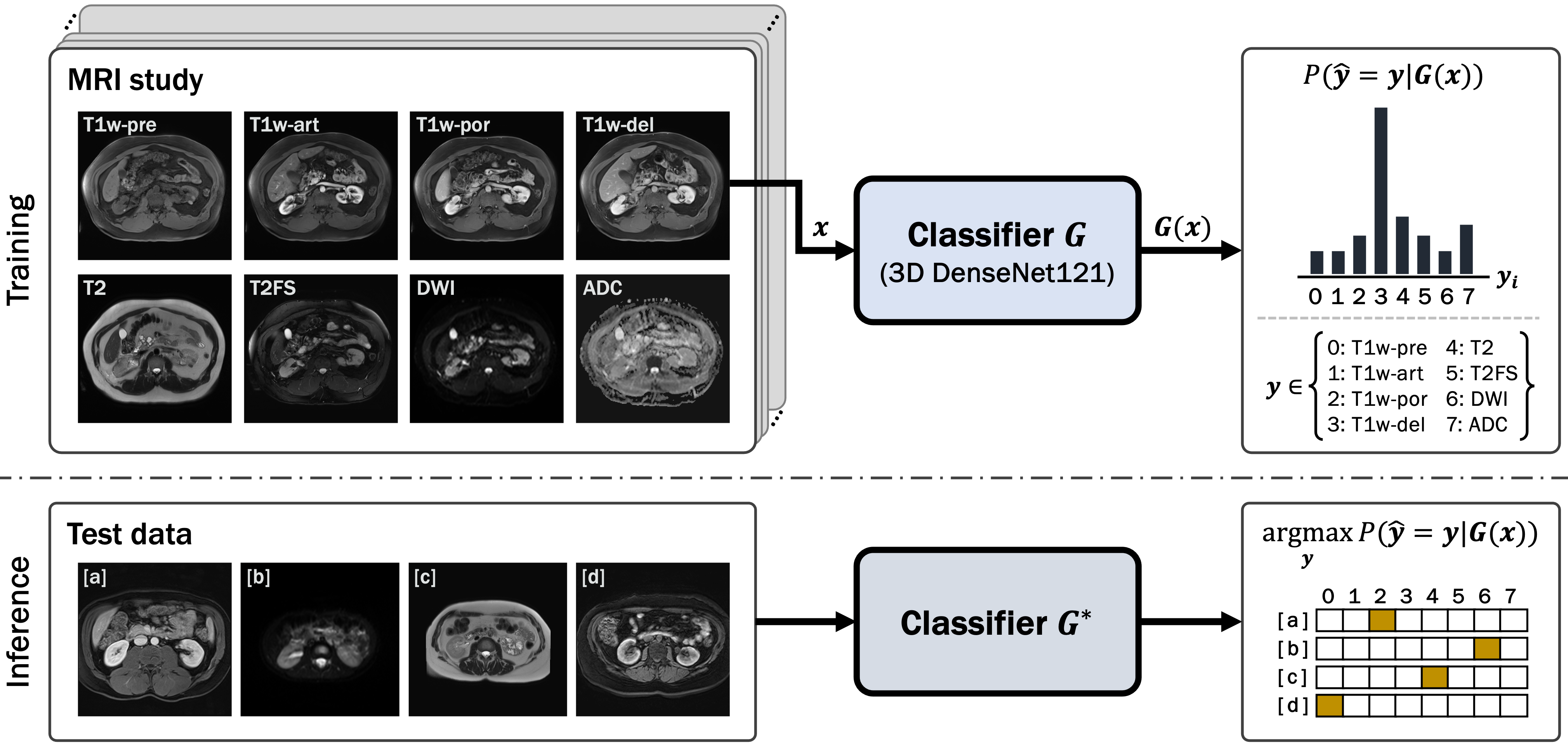}
\caption{The overall framework of our body mpMRI series classifier. Eight different MRI series \{pre-contrast T1, T1-arterial phase, T1-venous phase, T1-delayed phase, T2-weighted, T2FS, DWI, and ADC\} were used to train a 3D DenseNet-121 model with 5-fold cross-validation. At test time, the ensemble of models from each fold classified the series type of an input 3D MRI volume in the held-out test set. }
\label{fig_money}
\end{figure*}

In this work, we develop a framework to automatically classify the various series in mpMRI studies acquired at the level of the chest, abdomen, and pelvis. Our pipeline is able to distinguish between 8 commonly acquired series in the MRI protocol, including pre-contrast T1-weighted, post-contrast T1-weighted (arterial, portal venous, delayed), T2-weighted, fat-suppressed T2 (T2FS), Diffusion Weighted Imaging (DWI), and Apparent Diffusion Coefficient (ADC) series. We trained a DenseNet-121 model with 5-fold cross-validation on data acquired by three different Siemens scanners. Our results demonstrate the superior performance of the ensemble on a held-out test dataset. To the best of our knowledge, we propose the first model to automatically classify the 8 series types in mpMRI studies acquired at the level of the chest, abdomen, and pelvis. Our pipeline covers $\sim$70\% of the acquired series in the body MRI protocols.

\section{Methods}

\subsection{Data}

Between January 2015 and September 2019, the PACS system at the NIH Clinical Center was queried for patients who underwent mpMRI studies of the chest, abdomen, and pelvis. A total of 2,231 studies were collected, and a majority of the patients (n = 1399) underwent imaging with three different Siemens scanners (Aera, Verio, and BioGraph mMR). The MRI studies for these patients were manually verified to contain all of the following 8 series: T1-weighted sequence scanned before the contrast agent injection (T1w-pre), three contrast-enhanced T1-weighted sequences in the arterial phase (T1w-art), portal venous phase (T1w-por), and delayed phase (T1w-del), T2-weighted (T2), fat-suppressed T2 (T2FS), diffusion-weighted imaging (DWI), and apparent diffusion coefficient (ADC) series that was computed from DWI. This process yielded a total of 1,676 studies from 1,216 patients. Table~\ref{table_datasetRegion} shows a breakdown of the MRI studies based on the body part examined.
%

For our work, we considered only series that were acquired in the axial plane and did not include localizers, coronal, or sagittal series in the dataset. Also, based on the observation that there were often multiple DWI sequences (minimum 1, maximum 3) acquired with low (0 - 200 $s/mm^2$), intermediate (400 - 800 $s/mm^2$), and high (800 - 1400 $s/mm^2$) b-values in each MRI study, we used all the available DWI sequences with different b-values in the studies. As shown in Table~\ref{table_datasetSplit}, the final dataset for each fold was split into training, validation, and testing sets at the patient level, such that there was no data leakage.

\begin{table}[!t]
\centering
\caption{The number of MRI studies for each body region. ``Abd'' denotes the abdomen region.}
\resizebox{\linewidth}{!}{
\begin{tabular}{l|ccccc}
\hline
\multicolumn{1}{c|}{\textbf{Region}} & Chest & Chest+Abd & Abd & Abd+Pelvis & Pelvis \\
\hline 
\textbf{\# studies} & 32 & 1 & 1,573 & 18 & 52 \\
\hline
\end{tabular}
}
\label{table_datasetRegion}
\end{table}

\begin{table}[!t]
\centering
\caption{The number of patients and MRI studies for each fold of training, validation, and test datasets.}
\resizebox{\linewidth}{!}{
\begin{tabular}{c|cc|cc|cc}
\hline
\multirow{2}{*}{\textbf{Fold}} & \multicolumn{2}{c|}{\textbf{Train}} & \multicolumn{2}{c|}{\textbf{Validation}} & \multicolumn{2}{c}{\textbf{Test}} \\
& \# patients & \# studies & \# patients & \# studies & \# patients & \# studies \\
\hline 
1      & 851  & 1202 & 122    & 161   & 243 & 313 \\
2      & 852  & 1192 & 121    & 171   & 243 & 313 \\
3      & 853  & 1185 & 120    & 178   & 243 & 313 \\
4      & 852  & 1201 & 121    & 162   & 243 & 313 \\
5      & 850  & 1199 & 123    & 164   & 243 & 313 \\  
\hline
\end{tabular}
}
\label{table_datasetSplit}
\end{table}

\subsection{Model}

An overview of our mpMRI classification pipeline is illustrated in Fig. \ref{fig_money}. In the training stage, a classifier $G$ took an MRI volume $x$ and was trained to maximize probabilities $P(\hat{y}=y|G(x))$, where $G(x)$ is the output of the classifier, $\hat{y}$ is the predicted sequence type, and $y$ is the ground-truth label corresponding to the input. For the objective function, we employed a cross-entropy loss that can be computed by:
\begin{align}
     \mathcal{L}(y, \hat{y}) = -\sum_i y_i \mathrm{log}(\hat{y}_i).
\end{align}
We trained the classifier networks with five-fold cross-validation, in which for each fold, the model weights associated with the epoch achieving the highest validation accuracy were saved. In the inference stage, the classification models with the fixed learnable parameters for each fold, $G^*$, were ensembled together by averaging the output of probabilities. Then, the final sequence type for the input test volume was predicted via the argmax function, i.e., $\mathrm{argmax}_y P(\hat{y}=y|G(x))$. 
 
For implementing the classification network $G$, similar to prior work \cite{Mello2021_ResNet18, Zhu2022}, we initially adopted the ResNet-50 neural network \cite{He_ResNet} for the classification of 3D multi-parametric MRI sequences. We also compared the results of this model against the DenseNet-121 architecture \cite{Huang2017_DenseNet121}, which previously described improved classification results over the ResNet models. These models were implemented using the Medical Open Network for Artificial Intelligence (MONAI) framework \cite{Cardoso_MONAI}. MONAI is an open-source machine-learning framework for medical imaging built on a PyTorch base. 

\section{Experiments \& Results}

\subsection{Implementation details}

For each MRI series volume, we resampled the data to have a consistent voxel size of $1.5 \times 1.5 \times 7.8$ mm across volumes. Also, the voxel intensities were normalized into the range of [1\%, 99\%] \cite{Kociolek2020} to standardize the intensity distribution. Next, we resized each volume into a spatial dimension of $256 \times 256 \times 36$ through center-cropping or zero-padding. In training, we randomly rotated the volumes with 90 degrees for data augmentation. 

Using a batch size of 2, the model was trained via the Adam optimization algorithm with a learning rate of .0001 for a total of 25 epochs. At inference time, we evaluated the model using the evaluation metrics of the precision, sensitivity, specificity, and $F_{1}$ scores. All experiments were conducted with the Pytorch library in Python using an NVIDIA DGX workstation running Ubuntu 20.04 LTS with a single NVIDIA A100-SXM4-40GB GPU.


\subsection{Results}

\begin{figure*}[!t]
\centering
\includegraphics[width=\linewidth]{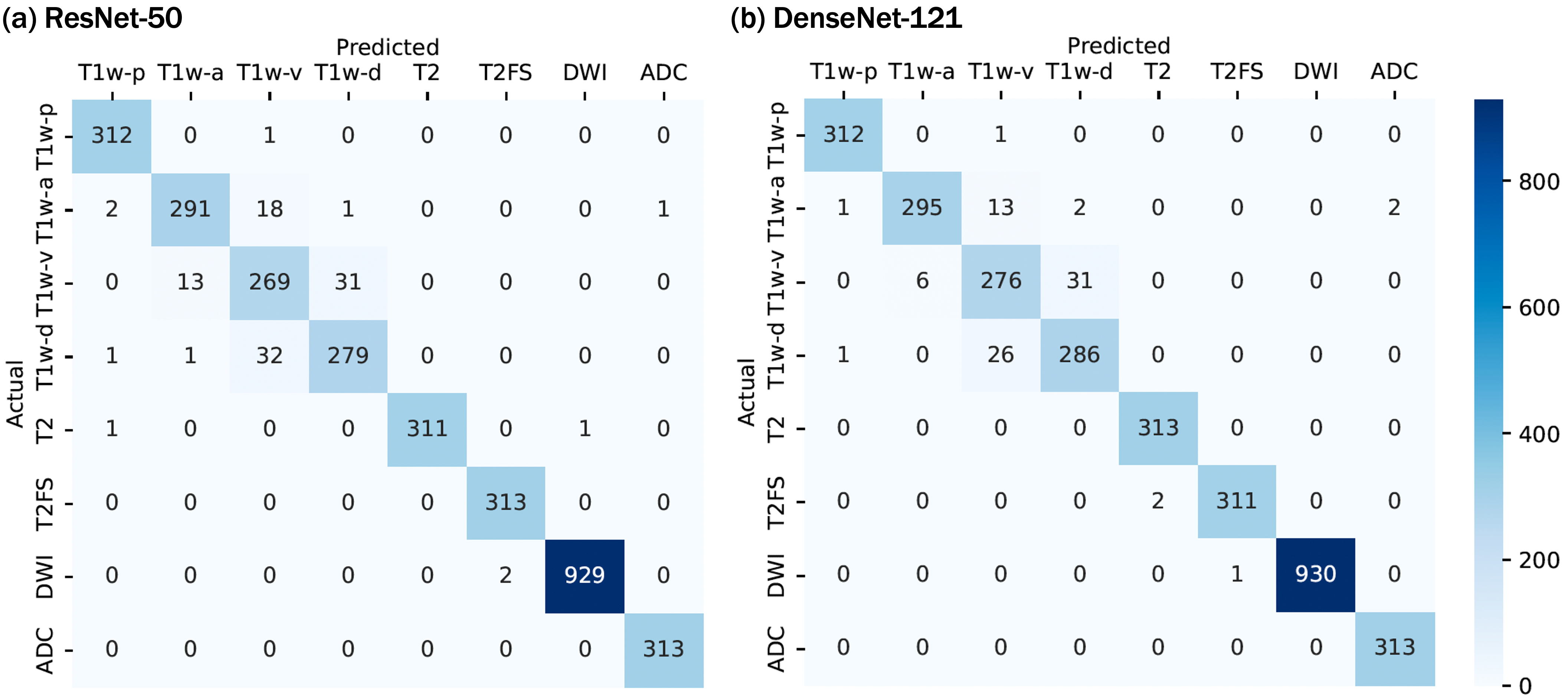}
\caption{Confusion matrix for our (a) ResNet-50 and (b) DenseNet-121 ensemble predictions on the test dataset (n = 313 studies). -p, -a, -v, and -d denote the pre-contrast, arterial, portal-venous, and delayed phase of T1-weighted imaging respectively.}
\label{fig_confusionMatrix_densenet}
\end{figure*}

We trained both the ResNet-50 and DenseNet-121 classifiers on the studies acquired by the Siemens scanners (n = 1363) and evaluated the performance of each model on the studies acquired by the same scanner in the test data split (n = 313). As shown in Table \ref{table_results_ensembleModelComparison}, we observed that the DenseNet-121 model had higher metrics across the board compared to the ResNet-50. The DenseNet-121 ensemble model achieved an overall average precision of 96.6\%, sensitivity of 96.6\%, specificity of 99.6\%, and $F_1$ score of 96.6\%. The confusion matrices for both classifiers are shown in Fig. \ref{fig_confusionMatrix_densenet}. These results indicated that the ensemble of the two classifiers correctly identified 5 out of 8 series with high precision. Table \ref{table_result_densenet} provides a breakdown of the metrics for the DenseNet-121 model for each series in the test data split. Although the T1 dynamic contrast-enhanced series (arterial, portal venous, and delayed phases) were confused most often with each other, the DenseNet-121 ensemble attained higher classification scores in contrast to the ResNet-50 ensemble. 




\begin{table}[!t]
\centering  
\caption{Comparison of classification results by the ResNet-50 and DenseNet-121 model ensemble on the test dataset (n = 313 studies). Bold font indicates best results.} 
\resizebox{\linewidth}{!}{
\begin{tabular}{l|cccc}
\hline
\multicolumn{1}{c|}{\textbf{Model}} & \textbf{Precision} & \textbf{Sensitivity} & \textbf{Specificity} & \textbf{F1 score} \\
\hline
ResNet-50 & 95.86 & 95.86 & 99.53 & 95.85 \\
DenseNet-121 & \textbf{96.60} & \textbf{96.59} & \textbf{99.62} & \textbf{96.59} \\
\hline 
\end{tabular}
}
\label{table_results_ensembleModelComparison}
\end{table}

\begin{table}[!t]
\centering  
\caption{Breakdown of classification results for DenseNet-121 model on our test dataset (n = 313 studies). The total score for each metric was computed by averaging the score across all series types.} 
\resizebox{\linewidth}{!}{
\begin{tabular}{l|cccc}
\hline
\multicolumn{1}{c|}{\textbf{Series}} & \textbf{Precision} & \textbf{Sensitivity} & \textbf{Specificity} & \textbf{F1 score} \\
\hline
T1w-pre & 99.36 & 99.68 & 99.93 & 99.52 \\
T1w-art & 98.01 & 94.25 & 99.79 & 96.09 \\
T1w-por & 87.34 & 88.18 & 98.58 & 87.76 \\
T1w-del & 89.66 & 91.37 & 98.83 & 90.51 \\
T2 & 99.37 & 100.00 & 99.93 & 99.68 \\
T2FS & 99.68 & 99.36 & 99.96 & 99.52 \\
DWI & 100.00 & 99.89 & 100.00 & 99.95 \\
ADC & 99.37 & 100.00 & 99.93 & 99.68 \\
\hline
\textbf{Total} & 96.60 & 96.59 & 99.62 & 96.59 \\
\hline 
\end{tabular}
}
\label{table_result_densenet}
\end{table}

\section{Discussion}


Our DenseNet-121 model achieved superior performance over the ResNet-50 model with an average specificity of 99.6\%, sensitivity of 96.6\%, and $F_1$ score of 96.6\%. It precisely classified ($\geq$ 95\%) pre-contrast T1, T2-weighted, T2FS, DWI and ADC series respectively. The confusion matrix in Fig. \ref{fig_confusionMatrix_densenet} also reflects this observation with fewer misclassifications. 

Despite this performance, several trends were observed. Primarily, the model had difficulty discerning between series that contained similar patterns in voxel intensity. For example, from Fig. \ref{fig_money}, the most common error was seen when the portal venous phase of T1-weighted imaging was incorrectly labeled as the delayed phase of T1-weighted imaging, and vice-versa. Similarly, the arterial phase of T1w imaging was also confused with the portal venous phase. However, these types of misclassifications are considered less significant or non-significant \cite{Zhu2022} by radiologists for the clinical display of the exam. These sequences are a constituent part of the post-contrast fat-suppressed T1-weighted series (same pulse MRI series type). The only difference between them is the elapsed time post-injection of the contrast material, and there are relatively inconsistent acquisition times across different institutions around the world.

A limitation of our work is the classification of only 8 series from the MRI protocols. With the consideration of the various sequences in this work, we have attempted to cover about 70\% of the MRI protocols at our institution. However, we have excluded a number of other series, such as the localizers, in-/out- phase T1w, sagittal, and coronal series in the mpMRI studies. Furthermore, depending on the anatomical region of interest, institutions around the world may have additional series that are a part of their MRI protocol. These include early arterial T1w imaging, late dynamic T1w imaging, inversion series, subtraction series etc. Some of these series fall into the same categories of series that we have chosen to classify in our work. For example: early arterial T1w series is often considered as the same series as arterial T1w, and late dynamic T1w is similar to portal-venous T1w \cite{Zhu2022}.

Another limitation of our work is the sole dependence on volumetric voxel data for the automated classification of the MRI series. Currently, we do not use the wealth of information from the DICOM headers that may prove beneficial for the classification task. Furthermore, the data used in this work was acquired from only one manufacturer (Siemens) at our institution. The generalization of the model ensemble performance to studies acquired by different manufacturers (e.g., Philips, GE, Toshiba) remains undetermined. A remedy for this issue would be to use MRI volumes from public datasets \cite{Macdonald2023} to maximize the variety of manufacturers in the training dataset.  



\section{ACKNOWLEDGEMENTS}      

\noindent
This work was supported by the Intramural Research Program of the NIH Clinical Center (project number 1Z01 CL040004). Boah Kim, Tejas Sudharshan Mathai, and Kimberly Helm have no conflicts of interest. Ronald M. Summers received royalties for patents or software licenses from iCAD, Philips, ScanMed, PingAn, Translation Holdings, and MGB, and his lab received research funding through a Cooperative Research and Development Agreement from PingAn.

\section{Compliance with Ethical Standards}

This study was approved by the Institutional Review Board (IRB) at the NIH and performed with retrospectively acquired patient data. The need for informed consent was waived.

\bibliographystyle{IEEEbib}

\end{document}